\documentclass[letterpaper, 10 pt, conference]{IEEEtran}

\usepackage{tabularx}

\usepackage{algpseudocode}
\usepackage{algorithm}
\usepackage{algorithmicx}

\usepackage{subfigure}
\usepackage{lipsum} 
\usepackage{arydshln}
\usepackage[dvips]{color}
\usepackage{comment}
\usepackage{todonotes}
\usepackage{epsf}
\usepackage{epsfig}
\usepackage{times}
\usepackage{epsfig}
\usepackage{graphicx}
\usepackage{bbold}
\usepackage{mathtools}
\usepackage{mathrsfs}
\usepackage{amssymb}
\usepackage{pdfpages}
\usepackage{epstopdf}
\newfloat{algorithm}{t}{lop}

\usepackage{amsmath}
\usepackage{dsfont}
\usepackage{lettrine} 

\usepackage{amsmath,epsfig,amssymb,algorithm,algpseudocode,amsthm,cite,url}
\usepackage{caption}

\allowdisplaybreaks
\usepackage{csquotes}

\usepackage{verbatim}
\usepackage[english]{babel}
\usepackage{amsmath,amssymb}

\usepackage{multirow}
\usepackage{rotating}

\usepackage[justification=centering]{caption}
\usepackage{verbatim}

\begin{document}
	%
	 \title{When Machine Learning Meets Wireless Cellular Networks: Deployment, Challenges, and Applications}
	\IEEEoverridecommandlockouts
	\author{\IEEEauthorblockN{Ursula Challita, Henrik Ryden, Hugo Tullberg\\}
			Ericsson Research, Stockholm, Sweden\\
           Email: \{ursula.challita, henrik.a.ryden, hugo.tullberg\}@ericsson.com


	}
	\maketitle
	
\IEEEpeerreviewmaketitle
	
\begin{abstract}
Artificial intelligence (AI) powered wireless networks promise to revolutionize the conventional operation and structure of current networks from network design to infrastructure management, cost reduction, and user performance improvement. Empowering future networks with AI functionalities will enable a shift from reactive/incident driven operations to proactive/data-driven operations. This paper provides an overview on the integration of AI functionalities in 5G and beyond networks. Key factors for successful AI integration such as data, security, and explainable AI are highlighted. We also summarize the various types of network intelligence as well as machine learning based air interface in future networks. Use case examples for the application of AI to the wireless domain are then summarized. We highlight on applications to the physical layer, mobility management, wireless security, and localization.
\end{abstract}

\section{Introduction}
Evolution to the 5th generation cellular technology (5G) and beyond networks will see an increase in network complexity - from new use cases to network function virtualization, large volumes of data, and different service classes such as ultra reliable low latency communications (URLLC), massive machine type communications (mMTC), and enhanced mobile broadband (eMBB). The latest standard, new radio (NR), was designed to be flexible in order to meet the new service requirements. Nevertheless, the increased flexibility implies a growing number of control parameters and an increased complexity which is forcing a fundamental change in network operations. Meanwhile, the recent advances in artificial intelligence (AI) promise to address the emerging complex communication system design. AI promises to combine simplification with improved performance and spectral efficiency and can therefore be regarded as a key component for increasing the value of 5G and beyond networks, if properly integrated into the system. 

At a technical level, AI will have a significant role in shaping future wireless cellular networks - from AI-based service deployment to policy control, resource management, monitoring, and prediction. Evolution to AI-powered wireless networks is triggered by the improved processing and computational power, access to massive amount of data, and enhanced software techniques thus enabling an intelligent radio access network and the spread of massive AI devices. Integrating AI functionalities in future networks will allow such networks to dynamically adapt to the changing network context in real-time enabling autonomous and self-adaptive operations. Network devices can implement both reactive and proactive approaches, where data driven algorithms would only replace or complement traditional design algorithms if there is an overall performance gain. In essence, AI techniques can be used to augment existing functions by providing useful predictions as input, replace a rule-based algorithm, and optimize a sequence of decisions such as radio resource management and mobility.

Existing literature have investigated the application of machine learning (ML) techniques to wireless networking. The authors in~\cite{mingzhe, MLair, tutorial_1} describe different applications of ML to wireless communication problems such as radio resource management and channel estimation. Nevertheless, such papers do not investigate deployment issues and network design challenges for aligning ML techniques to applications in wireless networks. The main scope of this paper is to summarize some of the main requirements needed for efficiently integrating AI functionalities in real networks while also highlighting on the benefits that AI brings to various wireless applications.

The main contribution of this paper is to review the role of AI in wireless networks. First, we summarize some of the key factors for successfully deploying AI functionalities in future networks. We discuss the distribution of network intelligence - in the device, base stations, and cloud - and elaborate on ML-based air interface. Second, we highlight on various components, such as secure data exchange and confidential computing, that are crucial for the successful integration of AI functionalities in cellular networks. Finally, we present various applications of AI to the wireless domain such as AI-based mobility and AI-assisted localization.

The rest of this paper is organized as follows. Section~\ref{sec:endTOend} elaborates on the distribution of network intelligence and ML-based air-interface. Section~\ref{sec:factors} provides a summary of the key factors for the successful integration of AI tools in future networks. Use case examples for the application of AI in wireless networking are summarized in Section~\ref{sec:applications}. Finally, conclusions are drawn in Section~\ref{sec:conclusion}.

%
%



\section{Key Factors for Successful AI Deployment}\label{sec:endTOend}
Future wireless networks must support flexible, programmable data pipelines for the volume, velocity, and variety of real-time data and algorithms capable of real-time decision making. Future communication networks must be designed to support exchange of data, models, and insights, and it is the responsibility of the AI agents to include any necessary user data. In this section, we provide an overview on the distribution of network intelligence and ML-based air interface, which are key components for realizing such vision in future networks.

\subsection{Distribution of Network Intelligence}
Future wireless networks will integrate intelligent functions across several layers of the network such as the wireless infrastructure, cloud, and end-user devices with the lower-layer learning agents targeting local optimization functions while higher-level cognitive agents pursuing global objectives and system-wide awareness. Such intelligent distribution can be categorized into three main types, namely autonomous node-level AI, localized AI, and global AI.

\begin{itemize}
  \item Autonomous node-level AI is used to solve self-contained problems at individual network components or devices, where no data is required to be passed through the network.
  \item Localized AI is where AI is applied to one network domain. It requires data to be passed in the network, however, is constrained to a single network domain, for example, radio access network or core network. Localized AI can also refer to scenarios where data is geographically localized.
  \item Global AI is where a centralized entity requires knowledge of the whole network and needs to collect data and knowledge from different network domains. Network slice management and network service assurance are examples of global AI.
\end{itemize}

\begin{table*}[t!]\footnotesize
\setlength{\belowcaptionskip}{0pt}
\setlength{\abovedisplayskip}{3pt}
\captionsetup{belowskip=0pt}
\newcommand{\tabincell}[2]{\begin{tabular}{@{}#1@{}}#1.6\end{tabular}}
 \setlength{\abovecaptionskip}{2pt}
 \renewcommand{\captionlabelfont}{\small}
 \captionsetup{justification=centering}
\caption{Benefits and challenges of global AI, localized AI, and autonomous node-level AI.}\label{AI_distribution}
\centering
\tabcolsep=0.03cm 
\begin{tabular}{|c|c|c|c|}
\hline
& Autonomous node-level AI & Localized AI & Global AI\\
\hline
Benefits & Ensures privacy & Data shared across different domains& Global optima \\
 &         Less delay & Favourable for power limited devices &\\
 &         No data overhead & & \\
 &         Reduced training complexity & & \\
\hline
Challenges &  Local optima & Security and privacy issues related to data sharing &Deployment issue \\
 & Device power and memory limitation & Data overhead & Data transfer cost \\
 & Computational limitation & Delay due to protocols and signalling & Training complexity \\
\hline
\end{tabular}
\vspace{-0.24cm}
\end{table*}
	
Table~\ref{AI_distribution} summarizes the benefits and challenges of autonomous node-level AI, localized AI, and global AI. Global AI in future networks can be deployed either per slice whereby each global AI entity manages a single slice independently of the other slices or per system whereby a single global AI manages all the network slices simultaneously. Another challenge here is the orchestration of different AI agents and the negotiation among them on the available radio resources. Meeting the requirements of all lower-level AI agents simultaneously can be challenging since this can result in resource shortage. Therefore, it is important to coordinate intelligence across different network domains in order to optimize the end-to-end network performance. This in turn has implications for system architecture - how to distribute the models and knowledge bases over the devices, base stations, and cloud (centralized versus distributed learning); whether model training should be offline or online; how to represent and prepare data for fast consumption by algorithms; and short-time scale versus long-time scale applications.

Centralized AI schemes can be challenging for some wireless communication applications due to the privacy of some features such as user location, limited bandwidth, and energy availability for data transmission for training and inference. This in turn necessitates new communication-efficient training algorithms over wireless links while making real-time and reliable inferences at the network edge. Here, distributed machine learning techniques (i.e., federated learning~\cite{federated_learning}) have the potential to provide enhanced user privacy and energy consumption. Such schemes enable network devices to learn global data patterns from multiple devices without having access to the whole data. This is realized by learning local models based on local data, sending the local models to a centralized cloud, averaging them and sending back the average model to all devices. Nevertheless, the effectiveness of such schemes in real networks should be further studied considering the limitations of processing power and memory of edge devices. As such, configurations for centralized, distributed, and hybrid architectural approaches should be supported. Moreover, it is vital to design a common distributed and decentralized paradigm to make the best use of local and global data and models.



\subsection{ML-based Air Interface}
In addition to the distribution of network intelligence, future wireless networks might comprise a fully end-to-end machine learning air-interface. Here, the challenge is to train an interface that can support efficient data transmission and reduce energy consumption while also fulfilling the latency requirements of different applications. While an ML air-interface can be trained for optimizing data transmission, it can be challenging to handle typical control channel problems such as energy efficiency in scenarios where no data is transmitted or received.
Moreover, an ML-based air interface system should be able to adopt to the different requirements of each network slice in terms of data throughput, energy efficiency, and latency. As such, initial AI deployments can focus on an ML air-interface targeting data transmissions improvement only. This approach is similar to the first NR non-standalone deployments where NR was introduced for eMBB services while being aided by existing 4G infrastructure. In NR non-standalone, the network uses an NR carrier mainly for data-rate improvements, while the LTE carrier is used for control tasks such as mobility management and initial cell search.

\begin{figure}[!h]
\centering
\includegraphics[width=2.7in]{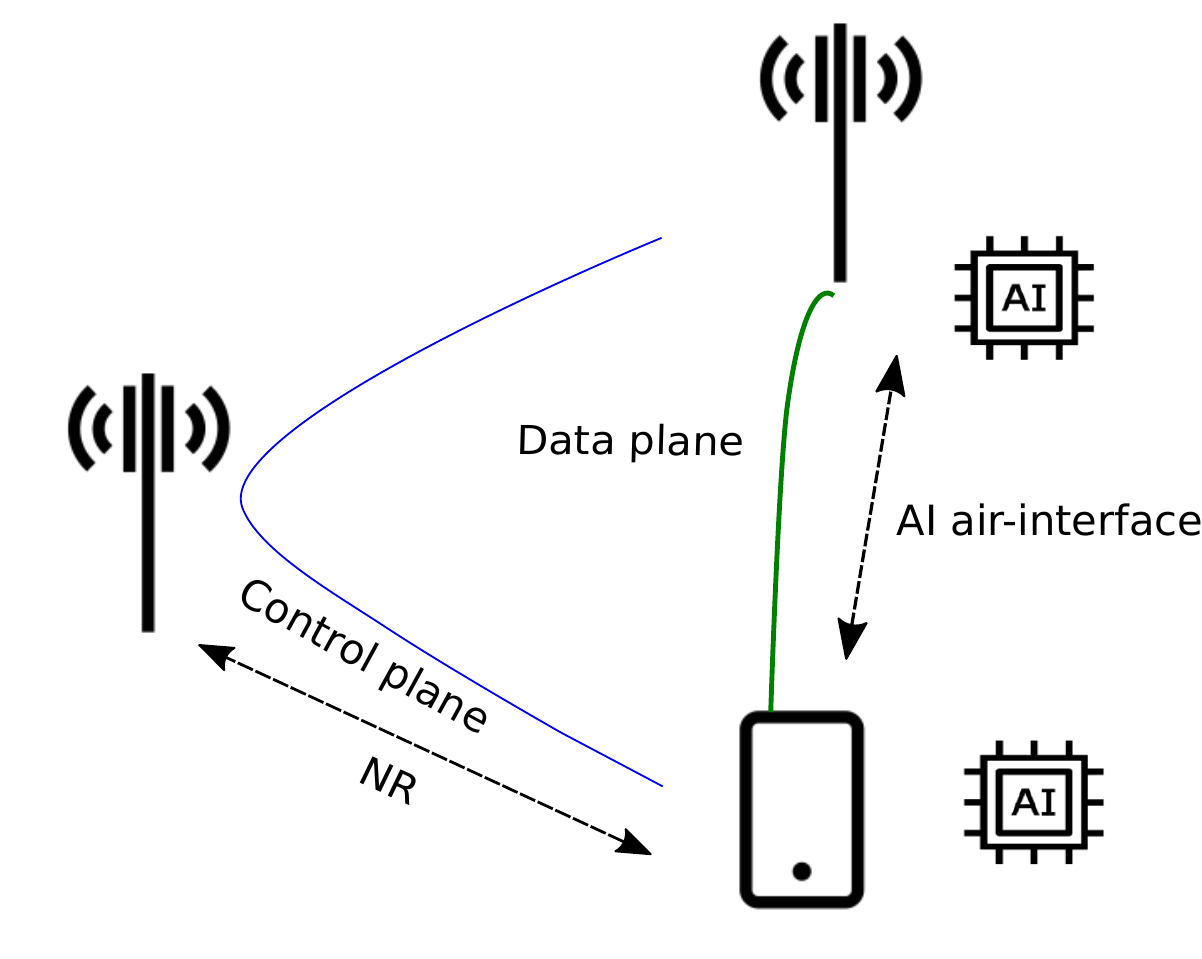}
\caption{Example of a potential future ML air-interface framework with an AI air-interface supported by a control plane on another radio access technology.}
\label{fig:ai_interface}
\end{figure}

A potential future ML air-interface, that is optimized for data-rate improvements, is illustrated in Fig. \ref{fig:ai_interface}. Using the primary radio access technology (RAT) (i.e., human-based designed RAT) for training the ML air-interface and for control signaling is an intermediate step towards the vision of a full AI-based RAT. Such signaling can detail how the users (UEs) should communicate on the ML-based air interface such as sending the structure and weights of a neural-network or indicating when/where the UE should receive data. Here, it is important to investigate the data volume for such transmissions and the length of the user-session for such a scheme to be beneficial.

Training the ML model could comprise of what bits the receiver should expect from the transmitter and what the receiver could feedback such as the loss. The network could update/train a potential auto-encoder based on the loss and feedback the updated weights to the transmitter/receiver~\cite{autoencoder}. The encoder part of the autoencoder would be at the transmitter side and the decoder at the receiver side.

In this section, we have highlighted on the deployment aspects of AI in future networks. To enable such deployment, next, we summarize some of the keys factors for the successful implementation of AI functionalities in mobile networks. 


\section{Key Factors for Successful AI Integration}\label{sec:factors}
AI tools must be tailored to the unique features and requirements of the wireless networks which are significantly different from the traditional applications of AI. In this section, we highlight some of the main areas that must be further investigated to realize the synergistic integration of AI in future wireless networks.

\subsection{Data acquisition} Acquiring and labelling data is fundamental when integrating AI in future wireless networks. The amount of data generated varies at different parts of the network. For example, physical-layer signals can generate gigabit of data every second from the BS, while data related to cell-level performance are generally much less. Here, domain competence is crucial in order to select the necessary data features, an efficient representation of the data, and the quality of the data, i.e., unbiased data sets. Moreover, the cost of signalling the data should be considered. For instance, data signalling from wireless devices is more costly than signalling over a wire (e.g., BS-BS signalling).

From an ethical perspective, the data acquisition process should account for the privacy of some of the features that can build the future AI networks. For example, knowing the geo-location of each device can improve network decisions such as paging, beamforming, and mobility decisions, however, it should only be considered with the consent of the device. To comply with the ethical considerations, several telecom industries follow the European commission on ethics guidelines for Trustworthy AI.

Moreover, several standardization organizations have proposed recommendations for the ML architecture to enhance the data acquisition process in future networks. For instance, the telecommunication standardization sector of ITU (ITU-T) proposes that the ML architecture must support data sharing between ML functionalities using distributed data sharing mechanisms and an interface to transfer data for training or testing models among ML functionalities of multiple levels\cite{ITU-T}.

\subsection{Data security and integrity} The success of integrating AI in future wireless networks will not only depend on the capability of the technology but also on the provided security. Data security related to wireless networks faces different challenges in comparison to traditional systems since it involves over-the-air information transmission which increases the risks of eavesdropping, false base stations, and jamming attacks. Building AI-models creates additional security threats whereby devices can intentionally report misleading training data. Therefore, it is important to guarantee building accurate data sets and AI models by avoiding data from malicious network devices. AI techniques can be adopted to prevent such attacks where, for instance, anomaly detection can be used to send an alarm when the system is compromised. 


Secure processing of the collected data is also crucial. A confidential computing multi-party data analytic can increase the user's and the network operator's trust in AI applications to the wireless network domain by ensuring that operators can be confident and that their confidential customer and proprietary data is not visible to other operators.

\subsection{AI implementation}
Future mobile networks must support a flexible AI implementation accounting for the performance of the AI application alongside resource allocation constraints such as resource availability, latency requirements, data availability, and training capabilities~\cite{ITU-T}. The model could, for instance, be executed at the device if the input model features are available at the device level or if signalling features is more costly than signalling the model to the device. Moreover, tight latency requirements can imply that the intelligence should be implemented closer to the device.


AI functionalities can be generally implemented at different network nodes (e.g., cloud, BS or user) and can be transferred across these nodes, an approach known as downloadable AI. The transferred model can include input features and model parameters such as neural network weights and structure. Here, model training, data and model storage, data and model compression, data and model transfer, data format, and online model update should be considered for the efficient implementation of AI algorithms in network devices. For instance, model update can be triggered by new quality of experience (QoE) metric such as the loss function or when the model output is above a threshold. The selection of the AI model should tailored to the device type to account for the difference in the memory limitations and computational capabilities of different devices.

Moreover, it is crucial to investigate model compression and acceleration techniques for signalling models over the wireless channel and for deploying them on devices~\cite{quantization}. For instance, deep learning techniques have received much attention in the wireless domain (see~\cite{DLsurvey}) due to their ability to handle highly dimensional state/action spaces. This is of particular interest to mobile networks due to the
increase in the number of BS antennas and the densification of the networks. Nevertheless, the downside
of existing deep neural network models is the memory requirement and energy consumption hindering their deployment
in devices with low memory resources such as IoT sensors.

\subsection{Robust and efficient AI learning}
AI systems need to be resilient and robust - they must be accurate, reliable, and ensure a fall back plan. Future mobile networks must support an ML sandbox for hosting separate ML pipelines to train, test, and evaluate ML functionalities before deploying them in a real environment. Here, the acceptance of some failures in the system can enable performing exploration in wireless networks. For instance, a block-error rate target of 10\% in the link-adaptation maps to an acceptance failure in every $10^{th}$ packet transmission. Exploration can be performed using reinforcement learning techniques~\cite{RL_book} while ensuring an acceptable failure rate. For instance, the level of exploration can be much lower or even zero in a critical communication setting whereas in mobile broadband settings the acceptance for short-term performance degradation can be higher. One can also identify network conditions for the underlying use case during which exploration can still guarantee the promised quality-of-service to the end-devices.

Robust AI can be also realized by initializing an AI model using conventional rule-based algorithms or in a realistic simulation or test bed environment where the AI agent can explore without limits and then have a principled way of transferring the knowledge to an operational real network. Transfer learning techniques can be used here for transferring the model to real environments. Given that mobile network environments often exhibit changing environment over time, transfer learning can also be adopted for transferring knowledge across different network domains such as in scenarios where the number of samples in the target domain is relatively small or if data becomes available at a relatively small time scale.

\subsection{AI goal alignment}
Introducing AI in wireless networks includes defining goals for such intelligence. Unlike other domains where the AI agent has a clear goal (e.g., winning GO and image classification), such goal is not always clear for wireless network applications. For instance, an AI-based scheduler that aims at maximizing the cell-level bitrate can achieve such goal by scheduling the users with best signal quality only. 

To address such limitation, AI alignment can be adopted in future wireless systems. AI alignment refers to the interaction of the AI agent with the user to take into consideration the user's goals and intentions during the learning phase. Such technique enables the AI model to align its behavior to the operator's or user's goals and is essential for scenarios where a built-in reward function is not available. The interaction between humans and machines can build trust and enable the machines to adjust their action to the human's intentions based on a suitable metric. A set of rules within which the AI can be aligned to the users desires but not cause general harm should be established.

\subsection{Active learning} Mobile networks are expected to support plugging in and out new data sources~\cite{ITU-T}. As mobile networks generate considerable amount of unlabeled data, data labeling becomes costly and requires domain-specific knowledge. Active learning schemes can be adopted for explicitly requesting labels to individual data samples from the user. Human-centered AI models is an example where the human is incorporated into the learning process enabling the AI system to learn from and collaborate with humans during the data annotation process~\cite{humancenteredAI}.


\subsection{Explainable AI techniques for cellular networks} Explainable AI refers to techniques where the outcome of the ML model can be explained/inferred and therefore strives a tradeoff between explainability and performance ~\cite{explainableAI}. The need for increased explainability to enable trust is crucial in future wireless systems due to the emerging wide range of mission and safety critical services such as V2X, remote surgery, and human machine brain interface. Incorporating explainable AI into future wireless systems can be realized, for instance, by developing an explainable twin AI system that can work in parallel to the AI system that is designed for performance optimization. Such an approach enables network operators to deploy performance-optimized AI systems while offering intuitive explanations to their corresponding decisions.

Finally, AI tools can be successfully incorporated into various wireless applications upon the realization of the aforementioned implementation challenges. Next, we summarize some of the main use cases of AI in mobile networks.


\section{Applications of AI in Wireless Networking}\label{sec:applications}
AI will inevitably be integrated at different levels of the network enabling operators to predict context information, adapt to the network changes, and proactively manage radio resources to achieve the network-level and user-level performance targets. On the short-term basis, applications of AI will mainly target separate network blocks such as scheduler and mobility management entity. On a long-term perspective, AI cross-layer design and optimization based on new QoE-based metrics is necessary for satisfying the end-to-end network performance requirements. To realize this, protocols need to be designed by violating the reference architecture enabling direct communication between protocols at non-adjacent layers, sharing variables, or joint tuning of parameters across different layers. Long-term AI applications also include the ML-based air interface as described in Section~\ref{sec:endTOend}.


Applications of AI techniques to the wireless network domain will essentially rely on various input features - radio-based features such as radio location and channel state information and non-radio features such as geographical location and weather conditions. For instance, the radio location comprises radio measurements on reference signals of the UE serving frequencies and is useful for different applications such as signal quality prediction, secondary carrier prediction, and beam alignment. Nevertheless, acquiring frequent UE measurements is costly and can result in signalling overhead which necessitates new efficient UE reporting formats and new report trigger events. It is also important to address the limitations of current AI algorithms such as the complexity, resource requirements, and energy consumption alongside the challenges summarized in section~\ref{sec:factors}. For instance, the energy consumed by the ML framework will be a tradeoff for using the available resources for increasing the user data rate.

Next, we elaborate on the application of ML techniques to different networking problems while highlighting on particular use cases.

\subsection{AI for Physical Layer}
\begin{figure}[t!]
	\begin{center}
		\centering
		\includegraphics[width=10cm]{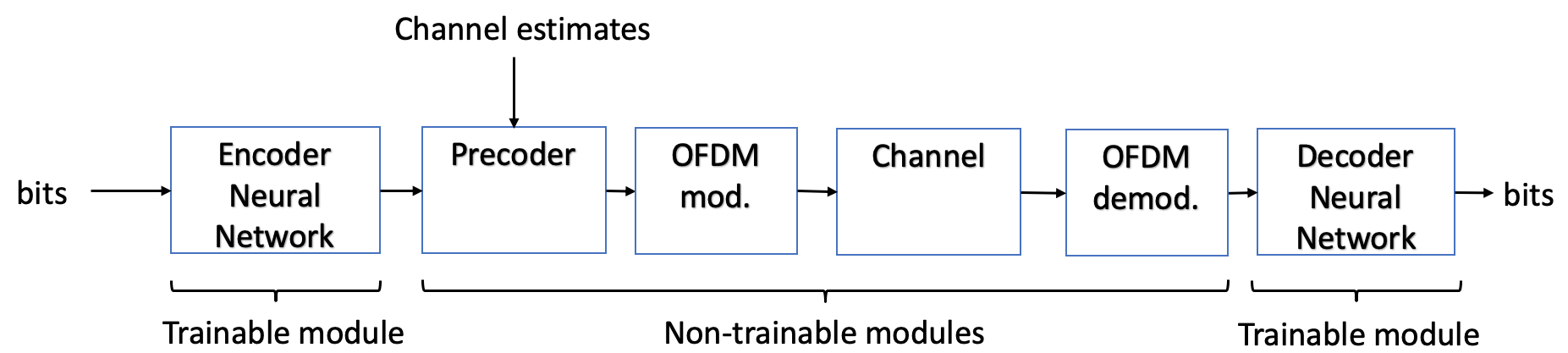}
		\caption{Example of a potential communication chain.}\label{fig:communication_chain}
	\end{center}
\end{figure}

The recent advancements in large steerable antenna arrays and cell-free architecture necessitates more coordination at the base stations~\cite{cell_free}. For example, forming the signal on each transmit antenna to maximize the signal quality at the UE side under imperfections such inter-node interference, channel estimation error, and antenna imperfections can be challenging and can be improved by ML techniques. AI techniques can also be adopted for improving separate modules in the transmission chain such as ML-based modulation.
Figure \ref{fig:communication_chain} provides an illustration of such a system highlighting on the trainable modules, the modulation layer, and the demodulation layer. Using this architecture, we exemplify the system with a single UE and a base station equipped with a large antenna array performing maximum ratio transmission precoding and assuming perfect channel estimates (estimated using e.g. UE sounding). One of the main challenges using an ML-based physical layer is to handle the channel distortion. By utilizing results from~\cite{hien}, we note that precoding the signal using MRT in a single antenna terminal deployment achieves a zero-phase with non-additive white gaussian noise channels. Although the phase can be zeroed due to channel hardening, the amplitude of the effective instantaneous channel remains unknown. 

\begin{figure}[t!]
	\begin{center}
		\centering
		\includegraphics[width=6cm]{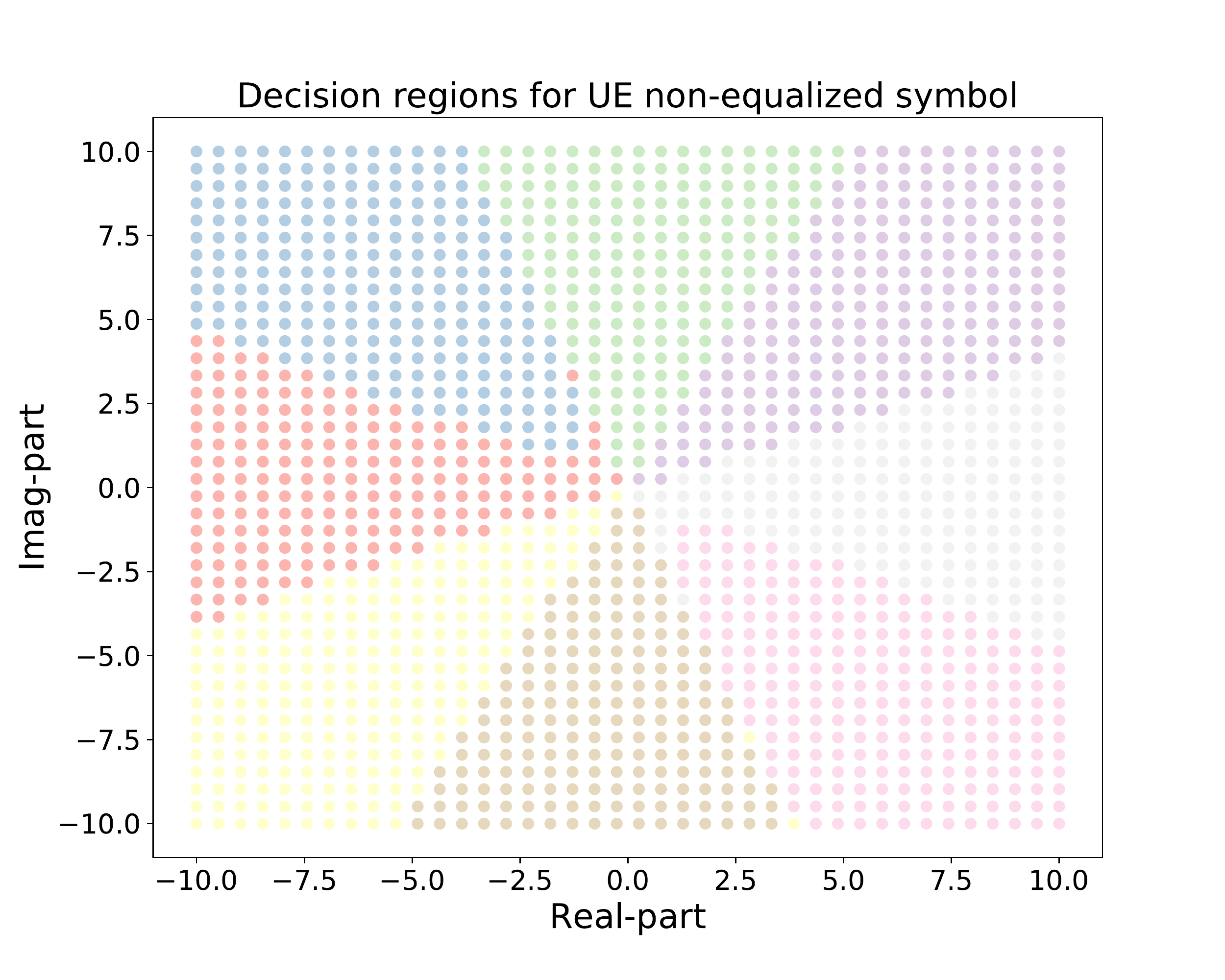}
		\caption{Decision regions for the trained 3-bit decoder (8 classes), axis comprise real and imaginary part of a received symbol in the frequency domain. The depicted received symbol region was arbitrary chosen between -10 and 10.}\label{fig:decision-region}
	\end{center}
\end{figure}

We consider an example where the network trains an auto-encoder comprising the modulation (encoder) and demodulation (decoder) and the base station intends to transmit 3 bits using a single carrier. 
The training can be performed using the concept described in Section \ref{sec:endTOend}.B, where the UE receives the intended bits on a primary carrier and feedback the loss and its neural network gradients on a secondary carrier. We evaluate the trained decoder network by feeding different non-equalized received symbols in the single carrier deployment into the decoder. The resulting class regions are illustrated in Figure \ref{fig:decision-region}. This figure shows how the trained decoder can split its received non-equalized symbols into different regions where each region comprises a unique class (8 possible classes for the considered 3 bit setup). Results show that the proposed ML framework can learn how to communicate with an unknown effective channel gain highlighting on the ability of the encoder/decoder to estimate the received bits using the phase-information of the received symbol.


\subsection{AI for Mobility Management}
Current wireless networks rely on reactive schemes for mobility management. However, such schemes might induce high latency that can be unfavorable for new emerging applications such as connected vehicles. Meanwhile, ML techniques enable proactive mobility decisions (see references [294]-[312] in~\cite{DLsurvey}), for instance, by predicting secondary carrier link quality, as described next.

NR will operate in 28 GHz and future wireless networks are likely to operate at even higher frequencies, increasing the need for tracking and prediction. Operation on 28 GHz leads to higher data rates and network capacity, however, it can result to less favorable propagation in comparison to lower frequencies, resulting in spotty coverage, at least in initial 28 GHz deployments. In order for the UEs to utilize also a potential spotty coverage on higher frequencies, the UEs need to be configured to perform inter-frequency measurements, which could lead to high measurement overhead at the device. An unnecessary inter-frequency measurement occurs when UEs are not able to detect any 28 GHz node, while not configuring a UE to perform inter-frequency measurements can result in under utilizing the large spectrum available at 28 GHz.

\begin{figure}[!h]
\centering
\includegraphics[width=3.7in]{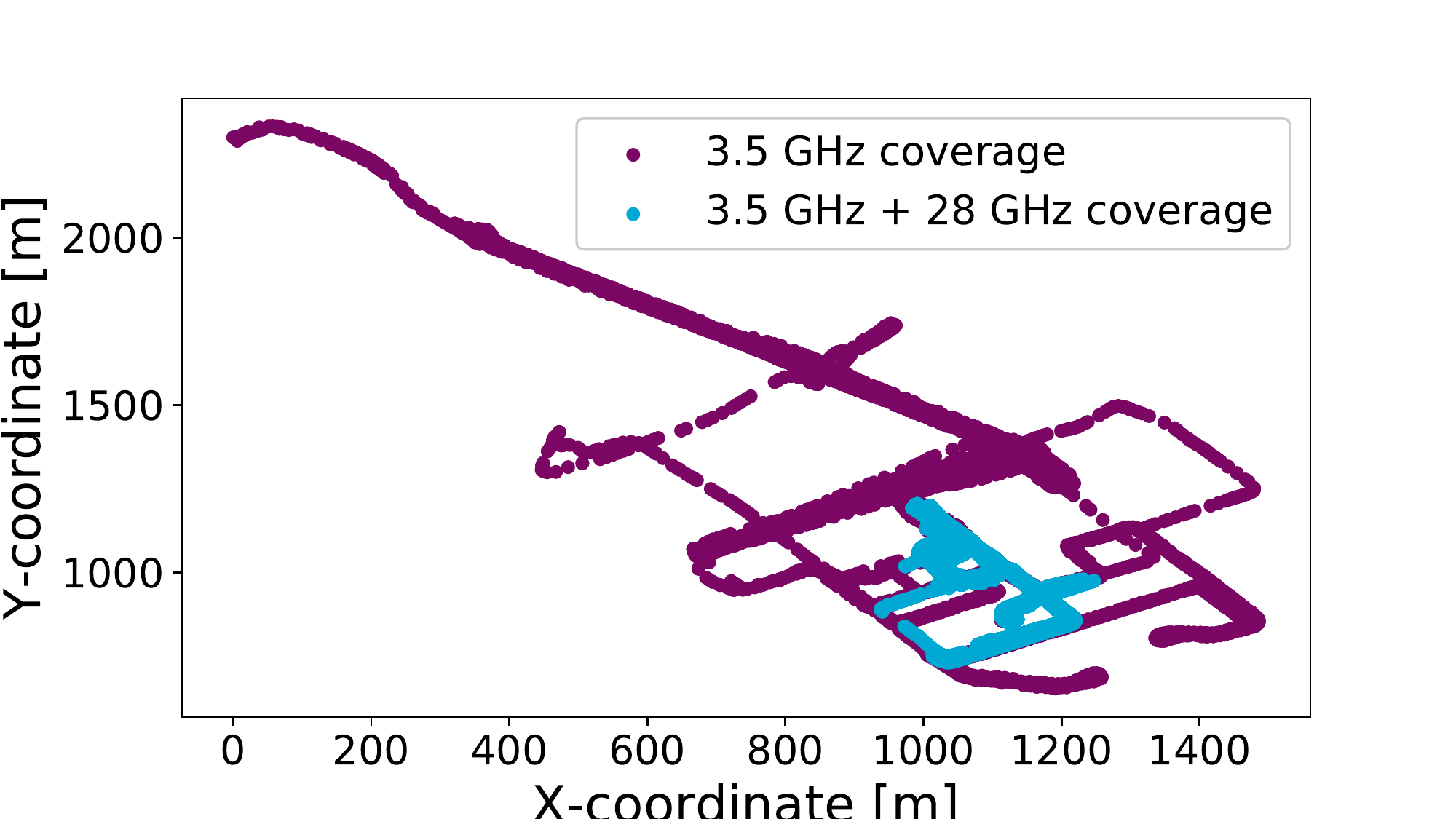}
        \begin{center}
        \text{(i)}
        \end{center}
\includegraphics[width=3.7in]{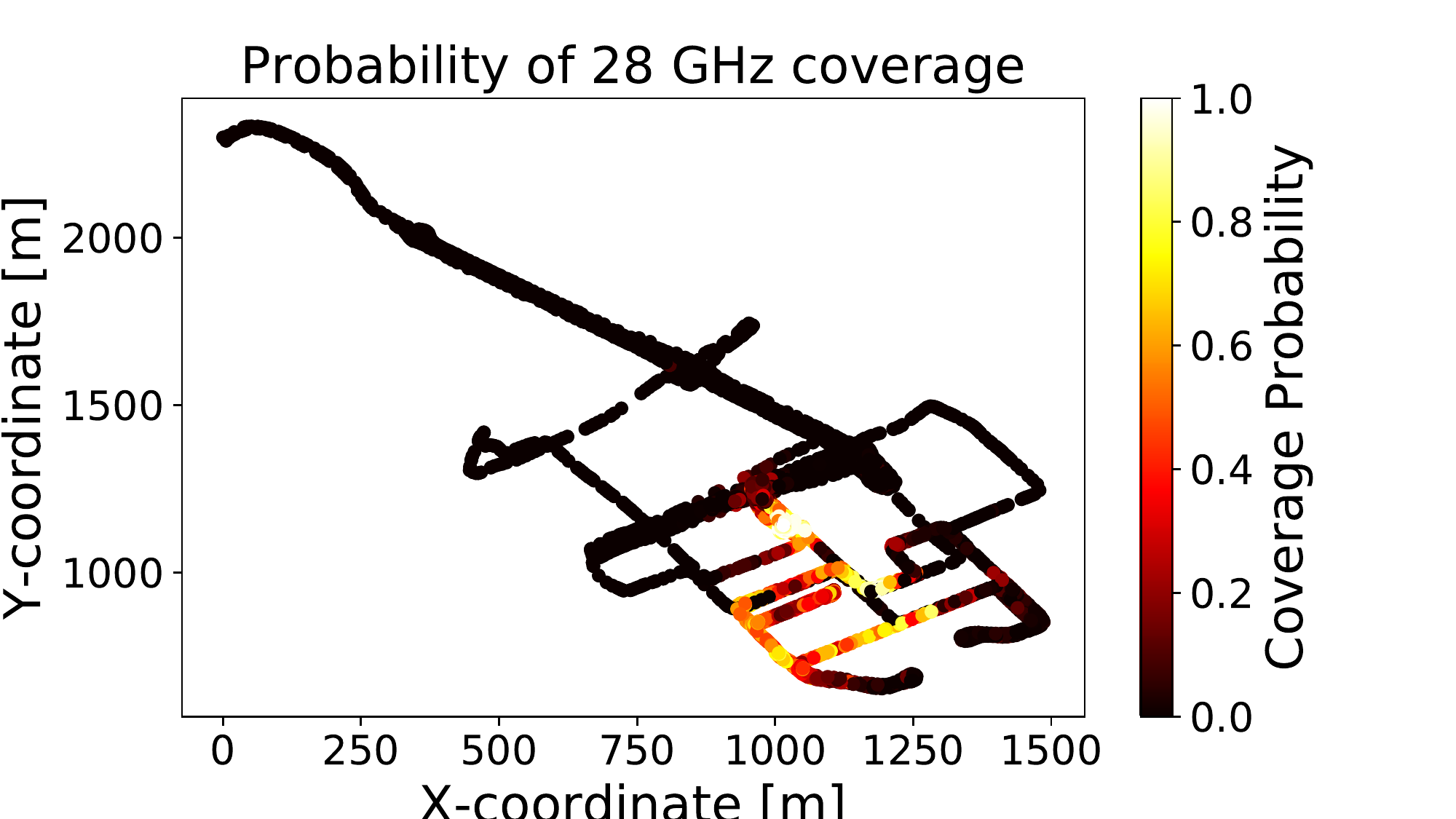}
        \begin{center}
        \text{(ii)}
        \end{center}
\caption{(i) Coverage on 28 GHz and 3.5 GHz over the evaluation area and (ii) predicted coverage probability on 28 GHz using measurements from a single 3.5 GHz node.}
\label{fig:prob28ghz_coverage}
\end{figure}

To limit the measurements on a secondary carrier, an ML scheme for predicting the coverage on the 28 GHz band based on measurements at its serving 3.5 GHz carrier node can be used. Figure \ref{fig:prob28ghz_coverage} (i) shows the coverage for the 3.5 GHz and 28 GHz for the scenario described in~\cite{3.5GHzScenario} i.e., based on UE measurements on both frequency bands. In this scenario, the 3.5 GHz node that is serving the UEs sweeps 48 beams and the UEs send reports for the beam strength of each beam.
Using secondary carrier prediction, Fig. \ref{fig:prob28ghz_coverage} (ii) shows the predicted coverage probability on the 28 GHz using measurements on 48 beams transmitted from a single 3.5 GHz node only, resulting in energy savings at the UE. With 36\% of the samples having coverage on the 28 GHz, an accuracy score of 87\% was achieved with a random forest classifier.



\subsection{AI for Wireless Security}
Maintaining high level of security for new use-cases and upon introducing AI in cellular networks is crucial for future wireless cellular networks. Alongside the data and model security issues mentioned earlier in Section~\ref{sec:factors}, ML techniques can be adopted for enhancing network security such as false base station identification, rogue drone detection, and network authentication (see references [406]-[421] in~\cite{DLsurvey}). For instance, detecting rogue cellular-connected drones is an important network feature since they can generate excessive interference to mobile networks disrupting their operation. Here, ML classification methods has been utilized for identifying rogue drones in mobile networks based on reported radio measurements (see ~\cite{rogue_drone}).


\subsection{AI for Localization}
New applications such as intelligent transportation, factory automation, and self-driving cars are important areas that drive the need for localization enhancements. The potentials with AI-based localization in wireless networks are expected to increase with massive antennas and new frequency bands in 5G which result in more unique radio-signal-characteristics for each location thus improving the localization accuracy using for example fingerprinting techniques (see references [313]-[336] in~\cite{DLsurvey}). For example, from Figure \ref{fig:prob28ghz_coverage}, one can note that there is only a subset of locations with coverage on both 3.5 GHz and 28 GHz resulting in a crude separation in geo-location based on radio measurements. Localization accuracy can be further improved by incorporating beam-measurements on different frequencies or by combining the received signal along with map information.


\section{Conclusion}\label{sec:conclusion}
In this paper, we provided an overview on the key factors for successful integration and deployment of AI functionalities in future cellular networks. We have presented use case examples for applications of AI to the radio access network while highlighting on the benefits that such techniques can bring to the network. An important research direction for future work is to investigate an ML-based architecture for end-to-end communication system design along with an ML-based air interface for initial deployments of AI-enabled wireless networks.

%
%

\bibliographystyle{IEEEtran}
\bibliography{references}

%
%

\end{document}